\documentclass{PoS}

\usepackage{amsfonts,amssymb,amsmath,bm}
\usepackage{graphicx}
\def\s#1{{\scriptscriptstyle #1}}

\def\gb{\bm{\Gamma}}
\def\l{{\bm{L}}}

\newcommand{\be}{\begin{equation}}
\newcommand{\bea}{\begin{eqnarray}}
\newcommand{\ee}{\end{equation}}
\newcommand{\eea}{\end{eqnarray}}

\title{Dynamical gluon mass generation and the IR sector of QCD}

\ShortTitle{Dynamical gluon mass generation and the IR sector of QCD}

\author{\speaker{Daniele Binosi}\\
        European Centre for Theoretical Studies in Nuclear Physics and Related Areas (ECT*),\\ Villa Tambosi, Strada delle Tabarelle 286, I-38050 Villazzano (TN), Italy\\
        E-mail: \email{binosi@ect.it}}

\abstract{We review the  Pinch Technique - Background Field Method (PT-BFM) framework for formulating and solving the Schwinger-Dyson equations of Yang-Mills theories. In particular, we show how within this framework it is possible to write a new set of Schwinger-Dyson equations that (i) accommodate the dynamical gluon mass generation through Schwinger's mechanism, and (ii) have  much better truncation properties than the the conventional equations. The resulting solutions show (in the Landau gauge) an infra-red saturating gluon propagator and ghost dressing function, in agreement with all lattice studies to date for both SU(2) and SU(3) gauge groups as well as 3 and 4 space-time dimensions. We also briefly discuss how a massive gluon enables self-consistently confinement through the condensation of thick vortices, and study other infra-red characteristic quantities such as the Kugo-Ojima function and the effective charge.}

\FullConference{Light Cone 2010 - LC2010\\
		June 14-18, 2010\\
		Valencia, Spain}

\begin{document}

\section{Motivation}

Even though the Green's (correlation) functions of pure Yang-Mills theories are not {\it per se} physical objects (for they show an explicit dependence on the gauge-fixing and renormalization scheme used), the understanding (or lack thereof) of their infra-red (IR) properties  has become an increasingly interesting topic in the last five years. The interest in these studies lies in the fact that 
reliable information on their non-perturbative structure is essential for unraveling the IR dynamics of QCD in general, and the mechanism behind confinement in particular. 

The exploration of the IR sector of such theories is currently pursued through mainly two non perturbative tools, namely the lattice -- where space-time is discretized and the quantities of interest are evaluated numerically -- and the Schwinger-Dyson equations (SDEs) -- the infinite set of integral equation governing the dynamics of the Green's functions. Each of the aforementioned methods has its own weaknesses, some of these being common to both (e.g., the need to fix a gauge, which in the lattice calculations generally coincides with the Landau gauge), some other being instead specific to the method under scrutiny. In the lattice case, for instance, one has to keep under control the various sources of systematic errors (e.g., discretization effects, finite volume effects, Gribov copies  effects), while at the same time providing enough computational power (that is large volume lattices) to study the deep IR region; in the SDEs case, the problem lies in devising a self-consistent truncation that would reduce the infinite tower of equations to a manageable subset,  
without introducing artifacts distorting the properties one has endeavored to investigate. Recently,   
a lot of progress has been made in addressing several of the aforementioned problems  in both methods~\cite{Cucchieri:2010xr,Aguilar:2006gr,Binosi:2007pi}, and we definitely have reached a point where SDE predictions can be systematically
compared against lattice results.

Over the last few years, in fact, high quality {\it ab-initio} lattice gauge theory computations have established beyond any reasonable doubt that the gluon propagator and the ghost dressing function of pure Yang-Mills in the Landau gauge saturates in the deep IR at a finite, non-vanishing value. The emergence of these {\it massive solutions} from the lattice together with their confirmation at the level of SDEs~\cite{Boucaud:2008ji,Aguilar:2008xm}, has caused a paradigmatic shift in our current understanding of QCD; in particular, the gluon mass generation scenario, put forth by Cornwall and others in the eighties~\cite{Cornwall:1981zr}, appears to be the preferred interpretation of what the underlying IR dynamics might be\footnote{The SDEs admit a second class of solutions that go under the name of {\it scaling solutions}. They are characterized by  an IR vanishing gluon propagator  and, correspondingly, an IR diverging ghost dressing function~\cite{Fischer:2006ub}, and have been predicted in confinement scenarios such as the one of Kugo-Ojima~\cite{Kugo:1979gm} and Gribov-Zwanziger~\cite{Gribov:1977wm}. So far, however, these solutions have not been found in (Landau gauge) lattice simulations.}. 
In this latter picture the gluon acquires dynamically an effective (momentum-dependent) mass, which accounts for the IR finiteness of the gluon propagator. The local gauge invariance and  (after gauge fixing) the BRST symmetry remain {\it intact}, because the original QCD Lagrangian is {\it not} altered at any point; indeed, the non-perturbative mechanism responsible for the mass generation is the four-dimensional generalization of the celebrated Schwinger mechanism~\cite{Schwinger:1962tn}.

\begin{figure}[!t]
\begin{center}
\includegraphics{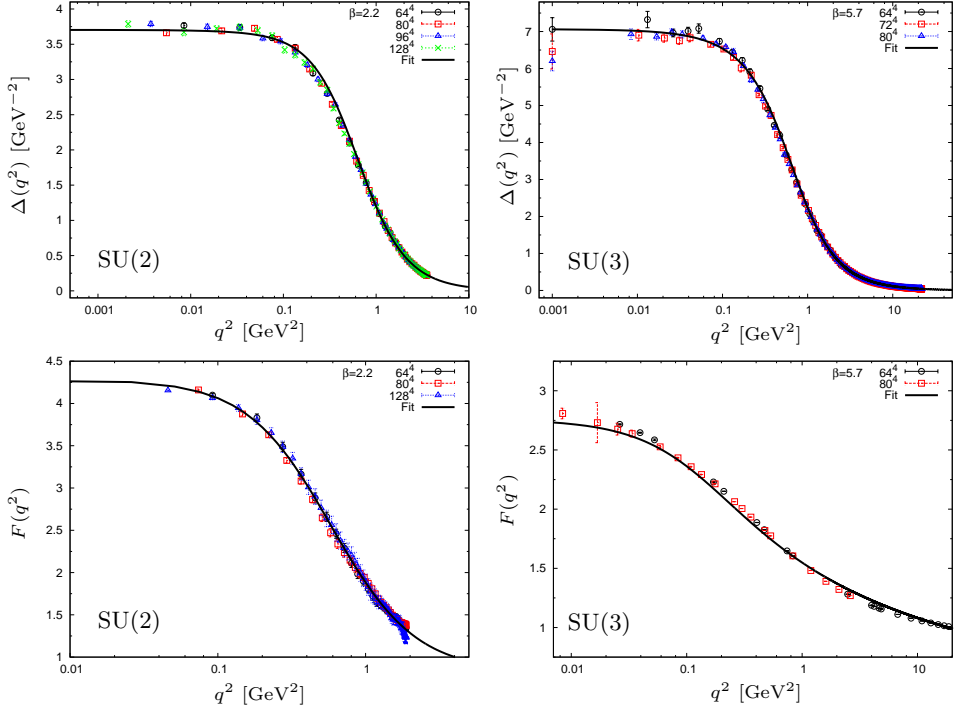}
\caption{Lattice data for the SU(2)~\cite{Cucchieri:2010xr,Cucchieri:2007md} and SU(3)~\cite{Bogolubsky:2007ud} gluon propagator and ghost dressing function. Solid lines corresponds to the best fit  given by the formulas (1.2) and (1.3).}
\label{fig1}
\end{center}
\end{figure}

In Fig.~\ref{fig1} we show a compilation of lattice data for the Landau gauge SU(2)~\cite{Cucchieri:2010xr,Cucchieri:2007md} and SU(3)~\cite{Bogolubsky:2007ud} gluon propagator (top panels) and ghost dressing function (bottom panels) from two independent groups. Although for each group the various simulation parameters vary (including lattice spacing, gauge group employed, and gauge fixing  algorithm), all results shares the common feature of the appearance of a plateau for the gluon propagator in the deep IR region, which is one of the most salient and distinctive predictions of the the gluon mass generation mechanism to be discussed below. In addition, the ghost dressing function show no enhancement in the deep IR; instead, it again saturates to a constant. 

Now, define the full gluon and ghost propagators in the $R_\xi$ gauges as
\be
\Delta_{\mu\nu}(q)=-\mathrm{i}\left[\Delta^{\mathrm{t}}_{\mu\nu}(q)+\xi\frac{q_\mu q_\nu}{q^2}\right],\qquad 
\Delta^{\mathrm{t}}_{\mu\nu}(q)=P_{\mu\nu}(q)\Delta(q^2), \qquad
D(q^2)=\mathrm{i}\frac{F(q^2)}{q^2}
\ee
where $P_{\mu\nu}(q)=g_{\mu\nu}(q)-q_\mu q_\nu/q^2$ is the transverse projector, and the scalar function $\Delta(q^2)$ is related to the gluon self-energy $\Pi_{\mu\nu}(q)=P_{\mu\nu}(q)\Pi(q^2)$ by $\Delta^{-1}(q^2)=q^2+\mathrm{i}\Pi(q^2)$; $F(q^2)$ above is the ghost dressing function, which is related to the ghost self-energy $\ell(q^2)$ through $q^2F^{-1}(q^2)=q^2-\mathrm{i}\ell(q^2)$. Given the above, one observes that the (Landau gauge) lattice data sets can be accurately fitted in terms of a massive gluon propagator and ghost dressing function of the type
\bea
\Delta^{-1}(q^2)&=& m^2(q^2) + q^2\left[1+ \frac{13C_{\rm A}g^2_1}{96\pi^2}
\ln\left(\frac{q^2 +\rho_1\,m^2(q^2)}{\mu^2}\right)\right],\nonumber \\
F(q^2)&=&1+ \frac94\frac{C_{\rm A}g^2_2}{48\pi^2}
\ln\left(\frac{q^2 +\rho_2\,m^2(q^2)}{\mu^2}\right),\nonumber \\
m^2(q^2)&=&\frac{m^4_0}{q^2+\rho m_0^2},
\label{fgluon}
\eea
where $C_A$ is the Casimir eigenvalue  of the gauge group's adjoint representation [$C_A=N$ for SU($N$)], while $\rho$, $\rho_{1,2}$ and $g_{1,2}$ are treated as free fitting parameters\footnote{For the SU(3) lattice simulations, $\mu$ must be chosen to coincide with the renormalization point, while for the SU(2) case it can be treated as an adjustable parameter.}. 
From the expressions above we clearly see the anticipated  dependence on the momentum transfer $q^2$ displayed by the dynamically generated mass; in particular, the latter assumes a non-zero value $m_0$ in the IR while dropping rapidly in the ultra-violet (UV) in a way consistent with the operator product expansion results~\cite{Lavelle:1991ve} (see next section).
Clearly $m_0$ acts as a physical mass scale: it regulates the perturbative renormalization group logarithm, 
so that, instead of diverging at the Landau pole, it saturates at a finite value. For the SU(3) case, the best fit is obtained for a gluon mass $m_0\sim 500$ MeV (with the data renormalized at $\mu=4.3$ GeV).

Lattice studies in the Landau gauge, therefore,  rule out the possibility of scaling solutions with nontrivial infrared exponents (and consequently the Kugo-Ojima scenario). In addition,  the original formulation of the Gribov-Zwanziger scenario is clearly disfavored, and must be drastically modified  through the inclusion of dimension two condensates, in order to be reconciled with the lattice data~\cite{Dudal:2008sp}.

In the rest of this talk we will concentrate on the  Pinch Technique - Background Field Method (PT-BFM)
framework~\cite{Cornwall:1981zr,Cornwall:1989gv,Binosi:2002ft}, where the aforementioned lattice findings 
may be naturally accommodated. Indeed, it should be noticed that the discovery of the key underlying ingredient, namely 
the dynamical   generation   of  a   gluon mass, coincided historically with the invention of the  PT~\cite{Cornwall:1981zr}, 
long before lattice simulations of QCD Green's functions were even contemplated.  

\section{Dynamical gluon mass generation, center vortices and confinement}

As mentioned earlier, already in the fifties  Schwinger~\cite{Schwinger:1962tn} pointed out that
the gauge invariance of a vector field does not necessarily 
imply zero mass for the associated particle.
Schwinger's idea is very simple.
Suppose that the vacuum polarization
 $\widetilde{\Pi}(q^2)$ [where $\Pi(q)=q^2\widetilde{\Pi}(q^2)$] 
acquires, for some  yet to be specified dynamical reason, a pole in the zero momentum transfer limit $q^2=0$, with positive 
residue $\mu^2$, i.e., $\widetilde{\Pi}(q^2) = \mu^2/q^2$.
Then (in Euclidean space)
$\Delta^{-1}(q^2) = q^2 + \mu^2$ and $\Delta^{-1}(0) = \mu^2$: that is, the 
vector meson has become massive, even if the gauge symmetry 
forbids a mass at the level of the fundamental Lagrangian.

There is {\it no} physical principle which would preclude $\widetilde{\Pi}(q^2)$ from 
acquiring such a pole, even in the absence of elementary scalar fields; 
in a {\it strongly-coupled} theory like non-perturbative Yang-Mills, for example,
this may happen for purely dynamical reasons, since strong binding may generate zero-mass bound-state excitations~\cite{Jackiw:1973tr}. The latter  act  {\it  like} dynamical Nambu-Goldstone bosons, in the sense that they are massless,
composite,  and {\it longitudinally   coupled};  but, at  the same  time, they
differ  from  Nambu-Goldstone  bosons   as  far  as  their  origin  is
concerned: they  do {\it not} originate from  the spontaneous breaking
of  any global symmetry\footnote{Notice that the 
usual Higgs mechanism corresponds to a particular case of Schwinger's mechanism, 
where the residue of the pole is saturated by the  
vacuum expectation value of an elementary scalar field.}~\cite{Cornwall:1981zr}. 

In the next section we shall see that the way the Schwinger mechanism is incorporated into the PT-BFM framework is through a particular form of the vertices appearing in the SDEs, which will display dynamical massless poles $\sim1/q^2$.  
The demonstration of the existence of such a zero-mass bound state, 
is a difficult dynamical problem, usually  studied by means of  Bethe-Salpeter equations
(see, e.g.,~\cite{Poggio:1974qs}). In what follows therefore we will rather {\it assume} that the theory can  
indeed generate the required poles, and see how far this assumption can lead us.

First of all, it should be noticed that a (gauge invariant) dynamically generated gluon mass cannot be a ``hard mass'', but will depend non-trivially on the momentum transfer $q^2$, and will decrease at large momentum (in order to avoid the existence of a corresponding bare mass). Indeed, application of the OPE within the PT framework~\cite{Lavelle:1991ve} shows that the gauge invariant contribution of the condensate  $\langle\mathrm{Tr}\ G^2_{\mu\nu}\rangle$ to the gluon propagator appears in the right way as to be interpreted as contributing to the running mass, with the asymptotic ultraviolet behavior $m^2(q^2)\sim\langle\mathrm{Tr}\ G^2_{\mu\nu}\rangle/q^2$ (actually, powers of logarithms of $q$ can also occur, but we ignore them here).
At this point, one can also write down an effective low-energy theory describing (gauge invariantly!) the low energy dynamics of massive gluons. To be precise, such a theory would be a ``quantum'' effective theory, since it resums all the quantum effects (not present in the classical action) that give rise to the dynamical gluon mass. The action we are looking for is the one corresponding to the  gauged non-linear sigma model and  known  as ``massive gauge-invariant Yang-Mills''~\cite{Cornwall:1979hz}, with    
Lagrangian density  
\begin{equation}
{\cal L}_{\s{\mathrm{MYM}}}= \frac{1}{2} F_{\mu\nu}^2 - 
m_0^2 {\rm Tr} \left[A_{\mu} - {g}^{-1} U(\theta)\partial_{\mu} U^{-1}(\theta) \right]^2.
\label{nlsm}
\end{equation}
In the formula above 
$A_{\mu}= \frac{1}{2\mathrm{i}}\sum_{a} \lambda_a A^{a}_{\mu}$, the $\lambda_a$ are the SU($N$) generators
(with  ${\rm Tr} \lambda_a  \lambda_b=2\delta_{ab}$), 
and the $N\times N$
unitary matrix $U(\theta) = \exp\left[\mathrm{i}\frac{1}{2}\lambda_a\theta^{a}\right]$ 
describes the scalar fields $\theta_a$.  
Note that ${\cal L}_{\s{\mathrm{MYM}}}$ is locally gauge-invariant under the combined gauge transformation 
\be
A^{\prime}_{\mu} = V A_{\mu} V^{-1} - {g}^{-1} \left[\partial_{\mu}V \right]V^{-1}, 
\qquad
U^\prime = U(\theta^\prime) = V U(\theta),
\label{gtransfb}
\ee
for any group matrix $V= \exp\left[\mathrm{i}\frac{1}{2}\lambda_a\omega^{a}(x)\right]$, where 
$\omega^{a}(x)$ are the group parameters. 

\begin{figure}[!t]
\begin{center}
%\hspace{.15cm}
\includegraphics[scale=0.75]{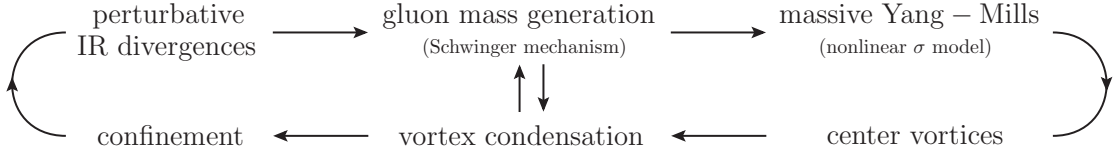}
\end{center}
\caption{\label{fig2.5}Self-consistent picture of confinement in the dynamical gluon mass generation scenario.}
\end{figure}

This action give rise to a plethora of possible ``quantum'' solitons, i.e.,  localized finite-energy configurations of gauge potentials corresponding to extrema of the effective action: center vortices, nexuses and sphalerons. Of these the most important are center vortices, since they show a  long-range pure  gauge term  in  their potentials,
which endows  them with a topological quantum  number corresponding to
the center  of the gauge group  [$Z_N$ for SU$(N)$]. This in turn is
responsible for quark  confinement and gluon screening. 
Specifically, center vortices of  thickness $\sim m_0^{-1}$,  form a condensate $\langle\mathrm{Tr}\ G^2_{\mu\nu}\rangle\neq0$ because their entropy
(per  unit  size) is  larger  than  their  action. This  condensation furnishes an  area law for the Wilson loop in  the fundamental representation, thus confining quarks~\cite{Cornwall:1981zr,Cornwall:1979hz}. 
In addition, the adjoint potential shows a roughly linear  regime followed by string breaking  when the potential
energy is about $2m_0$,  corresponding to gluon screening~\cite{Bernard:1982my}.

%\pagebreak

Thus, a dynamically generated gluon mass lead us to the confinement scenario of Fig.~\ref{fig2.5}

\section{Schwinger-Dyson equations in the PT-BFM}

At this point one would like to be more quantitative, and use the SDEs in order to thoroughly study the behavior of the gluon propagator and ghost dressing function in the dynamical gluon mass generation scenario. The key ingredient enabling such an analysis has been the development of a gauge invariant truncation scheme for non-Abelian SDEs~\cite{Binosi:2007pi}.

\begin{figure}[!t]
\begin{center}
\includegraphics[width=12cm]{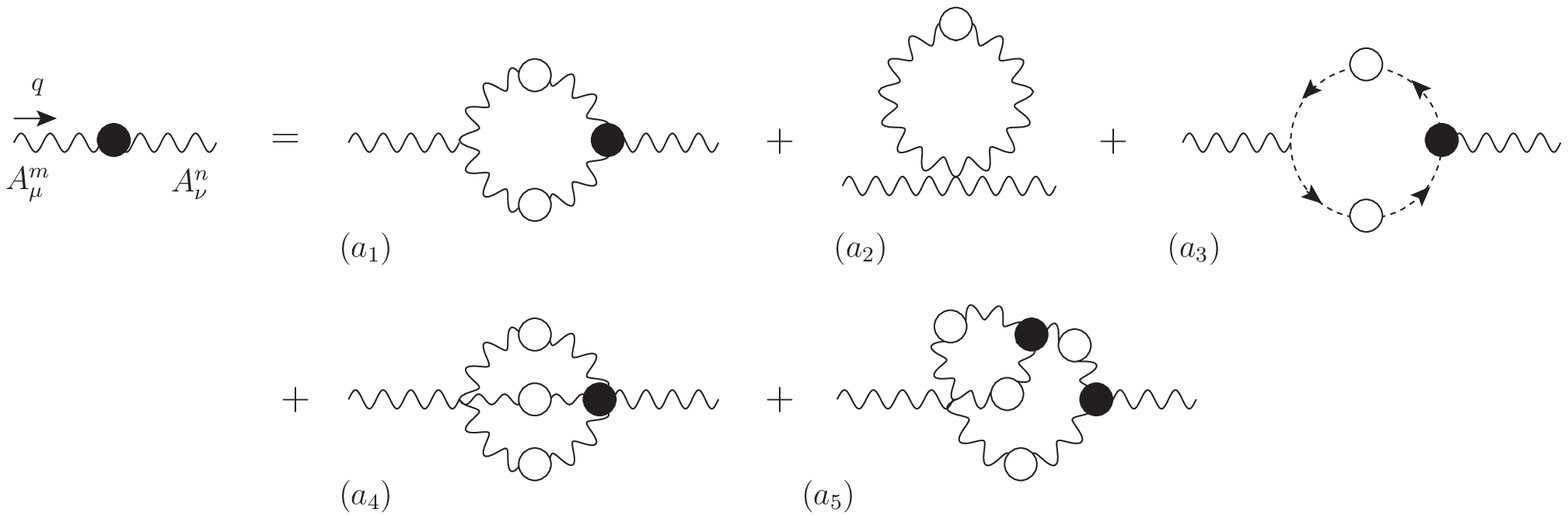}
\caption{SDE for the conventional gluon self-energy.}
\label{fig2}
\end{center}
\end{figure}

To understand where the problem lies, let us consider the conventional SDE for the gluon self-energy (Fig.~\ref{fig2}).
Due to the non-Abelian nature of Yang-Mills theories, it is highly non-trivial to verify diagrammatically the transversality of the gluon propagator, namely $q^\nu\Pi_{\mu\nu}(q)=0$.
%\be
%q^\mu\Pi_{\mu\nu}(q)=q^\mu\sum_{i=1}^5(a_i)_{\mu\nu}=0.
%\ee
The main reason for this resides in the nature of the Slavnov-Taylor identities satisfied by the vertices appearing in the SDEs. These always involve auxiliary ghost Green's functions, and are either extremely complicated (four-gluon vertex) or cannot be cast in a useful form (ghost-gluon vertex). As a result, it is hard to determine which set of diagrams can be safely discarded in the gluon SDE (or in other words how to {\it truncate} the SDE) without violating the transversality property (i.e., the gauge invariance) of the answer. Keeping only diagrams $(a_1)$ and $(a_2)$ is not correct even at one-loop, and adding $(a_3)$ will not improve the situation beyond one-loop.  

\begin{figure}[!t]
\begin{center}
\includegraphics[width=13cm]{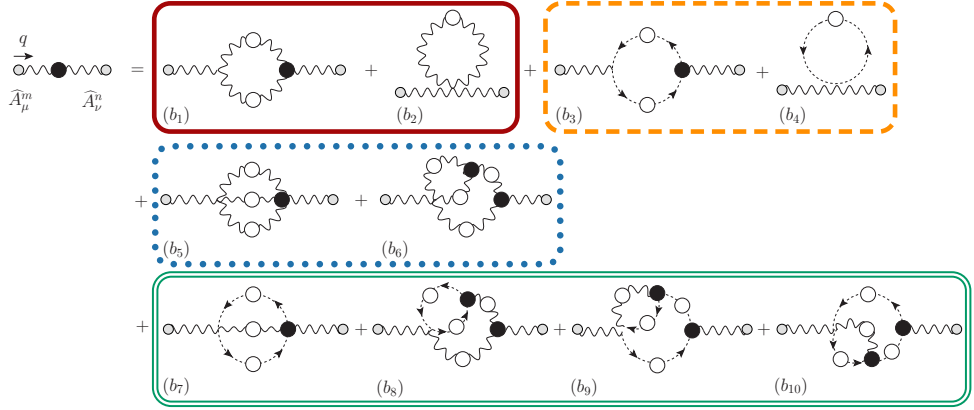}
\caption{The SDE for the gluon self-energy in the PT-BFM framework.}
\label{fig3}
\end{center}
\end{figure}

To overcome this limitation, one needs to devise a truncation scheme for the non-Abelian SDEs  that respects gauge invariance at every level of the {\it dressed-loop} expansion. This has been accomplished in~\cite{Binosi:2007pi} where the pinch technique has been used to carry out a systematic rearrangement of the entire Schwinger-Dyson series. For the case of the gluon self-energy, this procedure results in a new SDE describing a modified Green's function, namely the PT-BFM self-energy $\widehat{\Pi}_{\mu\nu}(q)$.
As shown in Fig.~\ref{fig3} this new SDE still contains the conventional gluon propagator but is composed by {\it new vertices} (indicated with a tilde in what follows) which corresponds precisely to the one to be found when quantizing the theory within the BFM: the external gluons have been projected dynamically into background gluons. This projection has far reaching consequences: when hit with the physical momentum $q$ the modified (BFM) vertices will now satisfy Abelian Ward identities, implying in turn that one- and two-loop gluon and ghost contributions are {\it individually} transverse~\cite{Aguilar:2006gr,Binosi:2007pi}.

\begin{figure}[!t]
\begin{center}
\includegraphics[width=8cm]{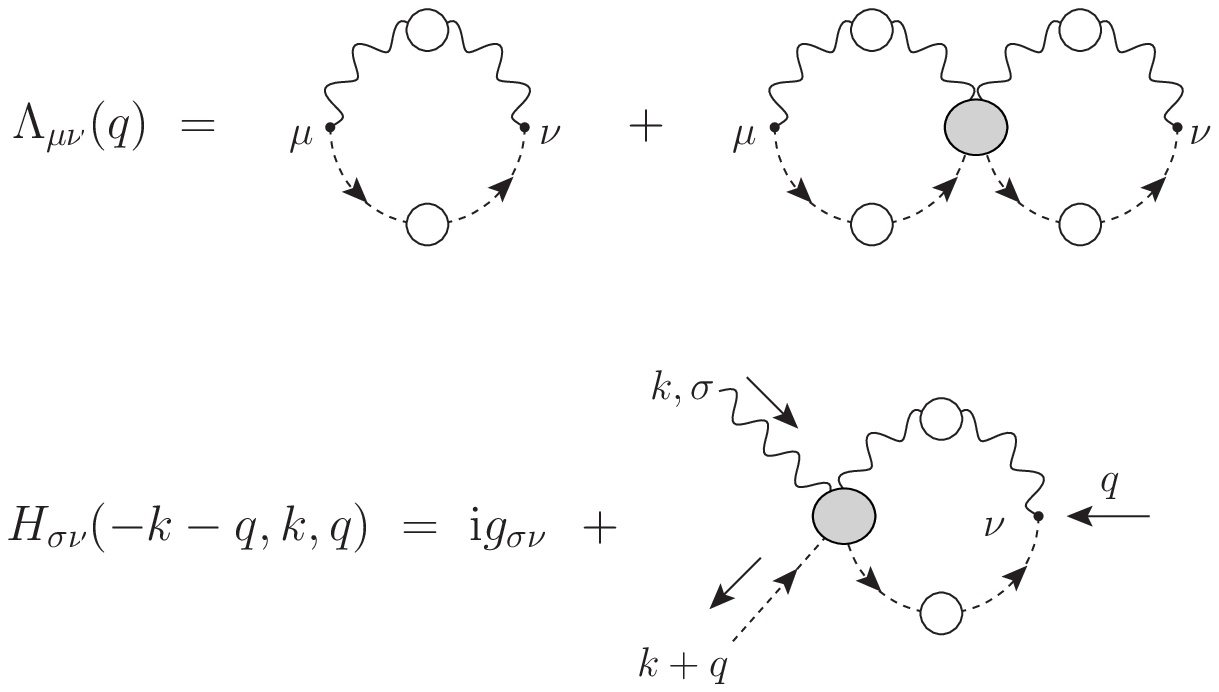}
\caption{Diagrammatic representation of the functions $\Lambda$ and $H$ appearing in the PT-BFM framework.}
\label{fig4}
\end{center}
\end{figure}

Notice that we do not have yet a dynamical equation for the conventional gluon propagator, since the quantity that appears at the lhs is not the one propagating in the diagrams of the rhs. This is solved by resorting to powerful identities known under the name of background quantum identities~\cite{Grassi:1999tp}, which relates $n$-point Green's functions with different content of background and quantum fields. In the case at hands one has
\be
\widehat{\Delta}^{-1}(q^2) =[1+G(q^2)]^2\Delta^{-1}(q^2), 
\label{BQI}
\ee
where $\widehat{\Delta}(q^2)$ is the background gluon propagator and $G(q^2)$ is the $g_{\mu\nu}$ form factor appearing in the Lorentz decomposition of the auxiliary Green's function 
$\Lambda_{\mu\nu}(q)$ (see Fig.~\ref{fig4})
\bea
\Lambda_{\mu \nu}(q) &=&g^2C_A
\int_k 
D(k+q)\Delta_\mu^\sigma(k)\, H_{\sigma\nu}(-k,-q,k,q)
\nonumber \\
&=&g_{\mu\nu} G(q^2) + \frac{q_{\mu}q_{\nu}}{q^2} L(q^2).
\label{LDec}
\eea
The function $H_{\mu\nu}$ (Fig.~\ref{fig4} again)
is in fact a familiar object, since it appears in the all-order Slavnov-Taylor identity
satisfied by the standard  three-gluon vertex. It is also related to the full gluon-ghost vertex $\gb_{\mu}$ by the identity
$q^\nu H_{\mu\nu}(-k-q,k,q)=-\mathrm{i}\gb_{\mu}(-k-q,k,q)$; at tree-level, $H_{\mu\nu}^{(0)} =\mathrm{i}g_{\mu\nu}$ and $\gb^{(0)}_{\mu}(-k,-q,k,q)=\Gamma_\mu(-k-q,k,q)=-q_\mu$. 

Considering only one-loop dressed diagrams (that is the only first two blocks in Fig.~\ref{fig3}) one has then the following system of integral equations\footnote{One should be aware of the fact that there is no a-priori guarantee that the one-loop dressed gauge-invariant subset kept  capture necessarily most of the dynamics, or, in other words, that they represent the numerically dominant contributions (however, for a variety of cases it seems to be true, see next section). But, the point is that one can systematically improve the picture by including more terms, without worrying that the initial approximation is plagued with artifacts, originating from the violation of gauge invariance.}
\bea
\Delta^{-1}(q^2)&=&\frac{q^2 +\mathrm{i}\sum_{i=1}^4(b_i)^\mu_\mu}{\left[1+G(q^2)\right]^2},\nonumber \\
\mathrm{i}D^{-1}(p^2)&=&p^2+\mathrm{i}\lambda\int_k\!\Gamma_{\mu}(-k-p,p,k)\Delta^{\mu\nu}(k)\gb_{\nu}(-k,-p,k+p)D(p+k),
\nonumber \\
G(q^2) &=& \frac{g^2 C_{\rm {A}}}{d-1}
\left[ 
\int_k \Delta^{\rho\sigma}(k)\, H_{\sigma\rho}(-k-q,k,q) D(k+q)
\right.\nonumber \\
&+&\left.\mathrm{i}
\frac{1}{q^2} \int_k\! q^{\rho} \Delta_{\rho\sigma}(k)\, {\gb}^{\sigma}(-k-q,k,q) D(k+q)
\right]\!\!,\qquad
\label{gSDE}
\eea
where in the first term the transverse projectors have been traced out by virtue of the block-wise transversality of the PT-BFM equation, and the identity (\ref{BQI}) have been used, to trade the BFM propagator for the conventional one.

Before attempting to solve the system, there is one last step to perform, that is the projection to the Landau gauge (since we want to compare the SDE results to lattice simulations). This is a subtle exercise in the one-loop dressed gluon sector, for one cannot set directly $\xi= 0$ in the integrals due to the terms proportional to $\xi^{-1}$ appearing in the BFM three-gluon vertex. Instead, one has to use the expressions for general $\xi$, carry out explicitly the set of cancellations produced when the terms proportional to $\xi$ generated by the	identity $k^\mu\Delta_{\mu\nu}(k)=-\mathrm{i}\xi k_\nu/k^2$ are used to cancel $1/\xi$ terms, and set $\xi=0$ only at the very end. It is relatively easy to establish that only the bare part of the full vertex contains terms that diverge as $\xi\to0$~\cite{Aguilar:2008xm}.
One has
\bea
\sum_{i=1}^2(b_i)^\mu_\mu&=&\frac{\lambda}{d-1}\Bigg\{\frac12\int_k\!\Gamma_\mu^{\alpha\beta}(q,k,-k-q)\Delta^{\mathrm{t}}_{\alpha\rho}(k)\Delta^{\mathrm{t}}_{\beta\sigma}(k+q)\l^{\mu\rho\sigma}(q,k,-k-q)\nonumber \\
&+&\int_k\!\Delta^{\mathrm{t}}_{\rho\mu}\frac{(k+q)_\sigma}{(k+q)^2}\left[\Gamma^{\mu\rho\sigma}(q,k,-k-q)+\l^{\mu\rho\sigma}(q,k,-k-q)\right]-(d-1)^2\int_k\!\Delta(k)\nonumber \\
&+&\int_k\!\frac{k^\mu(k+q)_\mu}{k^2(k+q)^2}\Bigg\},
\label{LGeq}
\eea
where $\l$ satisfies the Ward identity
\be
q^\mu\l_{\mu\rho\sigma}(q,k,-k-q)=\Delta^{-1}(k+q)P_{\rho\sigma}(k+q)-\Delta^{-1}(k)P_{\rho\sigma}(k).
\ee
For the ghost diagrams one finds finally
\be
\sum_{i=3}^4(b_i)^\mu_\mu=\lambda\int_k\!\widetilde{\Gamma}^\mu(-k-q,q,k)D(k)D(k+q)\widetilde{\gb}_\mu(-k,-q,k+q)-2\mathrm{i}d\lambda\int_k\!D(k).
\ee

\section{Approximations, numerical results and comparison with lattice data}

Before attempting to solve the PT-BFM equations derived in the previous section, one needs to identify suitable approximations for the various vertices appearing in them. As far as the conventional gluon-ghost vertex entering in the ghost equation, lattice studies (in $d=4$) indicate that it deviates very mildly from its tree-level value which will be therefore kept throughout (the same tree-level approximation will be used for the auxiliary vertex function $H_{\sigma\nu}$). 
Instead, the strategy one should adopt
for the background three-gluon and ghost-gluon vertices entering into the gluon propagator SDE, 
is the following. Given that the proposed truncation scheme relies crucially on the validity of the Ward identities satisfied by the background vertices,  one should start out with an approximation that manifestly preserve them. The way to enforce this, familiar to the SDE practitioners already from the QED era, is to resort to the {\it gauge-technique}~\cite{Salam:1963sa}, namely {\it solve} the Ward identities. Specifically, one must express the  vertices as a functional of the corresponding self-energies, in such a way that (by construction) the corresponding Ward identity is automatically satisfied\footnote{Notice that this procedure leaves the {\it transverse} (i.e., automatically conserved) part of the vertex undetermined. This is where the SDE for the vertex enters: It is used precisely to determine these transverse parts. Since however the transversality of the self-energy depends on  the {\it longitudinal} part of the vertex  only, one can apply all sort of approximations for the determination of the transverse form factors.}. One possibility, that contains the massless poles needed for the triggering of the Schwinger mechanism  is~\cite{Aguilar:2008xm}
\bea
\l_{\mu\rho\sigma}(q,k,-k-q)&=&\Gamma_{\mu\rho\sigma}(q,k,-k-q)+\mathrm{i}\frac{q_\mu}{q^2}\left[
\Pi_{\rho\sigma}(k+q)-\Pi_{\rho\sigma}(k)\right], \nonumber \\
\widetilde{\gb}_{\mu}(q,k,-k-q)&=&\widetilde{\Gamma}_\mu(q,k,-k-q)-\mathrm{i}\frac{q_\mu}{q^2}\left[\ell(k+q)-\ell(k)
\right].
\label{vtx}
\eea
Other forms are also allowed; in particular in~\cite{Aguilar:2009ke} a vertex that avoids the seagull divergences that the expressions~(\ref{vtx}) are bound to generate has been derived.

The methodology described above constitutes, in fact, the standard procedure even in the context of QED, where the structure of the SDE is much simpler (the SDE for the photon contains one single graph that involves the photon-electron vertex which satisfies automatically a Ward identity). Thus, while the PT-BFM approach described here replicates QED-like properties at the level of the SDEs of QCD, which is a striking fact in itself, does not make QCD easier to solve than QED.
One should appreciate an additional point though: any attempt to apply the approach described above in the context of the conventional SDEs is bound to lead to the violation of the transversality of the gluon self-energy, because (i) the vertices satisfy non-linear Slavnov-Taylor identities (a fact that makes the application of the gauge-technique impractical), and (ii) even if one had managed to come 
up with good Ans\"atze for all vertices, one should still keep all self-energy diagrams in Fig.~\ref{fig2} to guarantee that $q^\nu\Pi_{\mu\nu}(q)= 0$. From this point of view, the improvement of the PT-BFM approach over the standard formulation becomes evident.

\subsection{Gluon and ghost Green's functions}

\begin{figure}[!t]
\begin{center}
\includegraphics[scale=0.9]{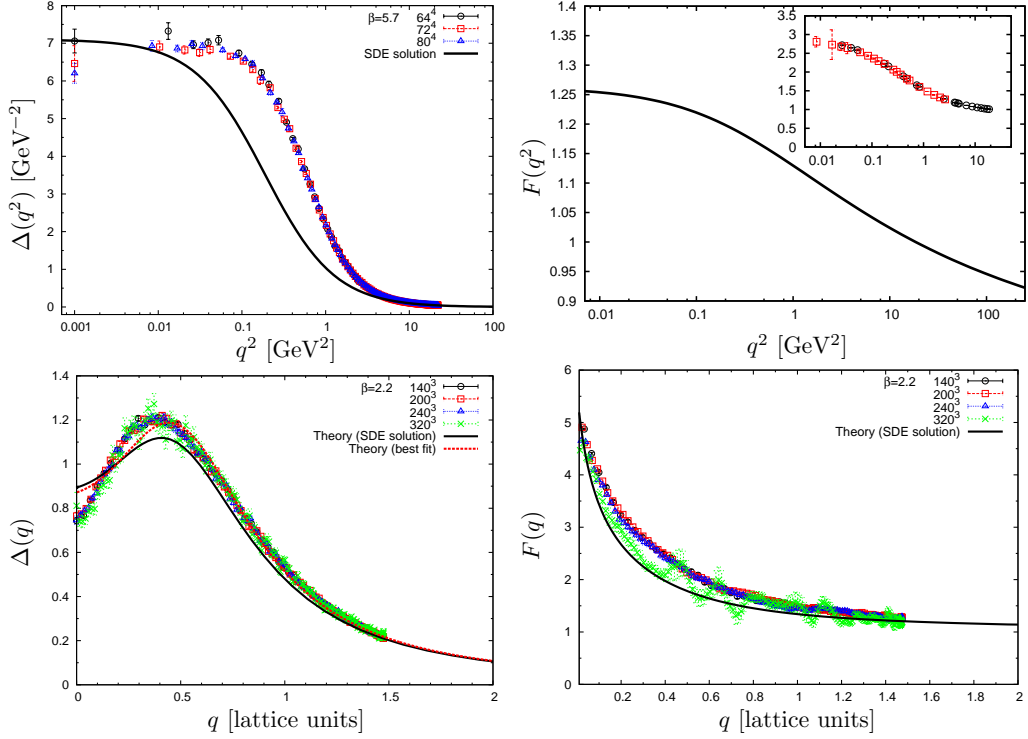}
\caption{Gluon propagator and ghost dressing function for the SU(3) $d=4$ and SU(2) $d=3$ cases, and comparison with the corresponding lattice data.}
\label{fig5}
\end{center}
\end{figure}

The solutions of the PT-BFM equations~(\ref{gSDE}) with the vertices~(\ref{vtx}) in the SU(3) $d=4$ case~\cite{Aguilar:2009ke}, and in the one-loop approximation for the SU(2) $d=3$ case~\cite{Aguilar:2010zx}, are shown in Fig.~\ref{fig5}.
As it can be seen in the plots, the PT-BFM equations capture qualitatively the correct $d=4$ physics even within the rough approximations used, while in the $d=3$ case the agreement becomes quantitative. In this latter case one gets the value $m/2g^2=0.153$ in agreement with many independent lattice studies.

The agreement found between the SDEs and the lattice results allows one to study other quantities of interest by using the lattice directly as an input into the various SDE. The general strategy adopted in this case is the following. One takes the lattice gluon propagator as an input for the ghost SDE; then solves for the ghost dressing function, tuning the coupling constant $g$ in such a way that the solution gives the best possible approximation to the lattice result. Obviously one must check that the coupling so obtained (at the renormalization scale used for the computation) is fully consistent with known perturbative results (obtained in the MOM scheme, which is the scheme used in our computations). Then the system is ``tuned'', and one can construct 
and analyze other quantities built out of $\Delta$, $F$ and $g$, as is done in the next two sections.

\subsection{The $G$ and $L$ auxiliary functions}

The first quantities that can be studied by means of the  procedure just described, are the auxiliary functions $G(q^2)$ and $L(q^2)$, which, within our approximations, are given by 
\bea
G(q^2) &=& \frac{g^2 C_{\rm {A}}}{d-1}\int_k \left[(d-2)+ \frac{(k \cdot q)^2}{k^2 q^2}\right]\Delta (k)  D(k+q),
\nonumber\\
L(q^2) &=& \frac{g^2 C_{\rm {A}}}{d-1}\int_k \left[1 - d \, \frac{(k \cdot q)^2}{k^2 q^2}\right]\Delta (k)  D(k+q).
\label{simple}
\eea 
Before doing that, let us notice that in the Landau gauge one can prove that the dressing function $F(q^2)$ and the form factor $G(q^2)$ and $L(q^2)$ are related through the BRST identity~\cite{Kugo:1995km}
\be
F^{-1}(q^2)=1+G(q^2)+L(q^2).
\label{id}
\ee
Since, under very general conditions on the gluon and ghost propagators, $L(q^2)\to0$ when $q^2\to0$ one has the IR relation $F^{-1}(0)=1+G(0)$. Thus, we see that a divergent (or {\it enhanced}) dressing function requires the condition $G(0)=-1$. The latter looks suspiciously similar to the Kugo-Ojima confinement criterion demanding that a certain function $u(q^2)$ (the Kugo-Ojima function) acquires the IR value $u(0)=-1$. Indeed it is possible to show that $G$ is nothing but the Kugo-Ojima function~\cite{Kugo:1995km}
\be
u(q^2)\equiv G(q^2),
\ee
and therefore $G(q^2)$ encodes practically all relevant information of the IR dynamics of the ghost sector. 

\begin{figure}[!t]
\begin{center}
\includegraphics[scale=0.93]{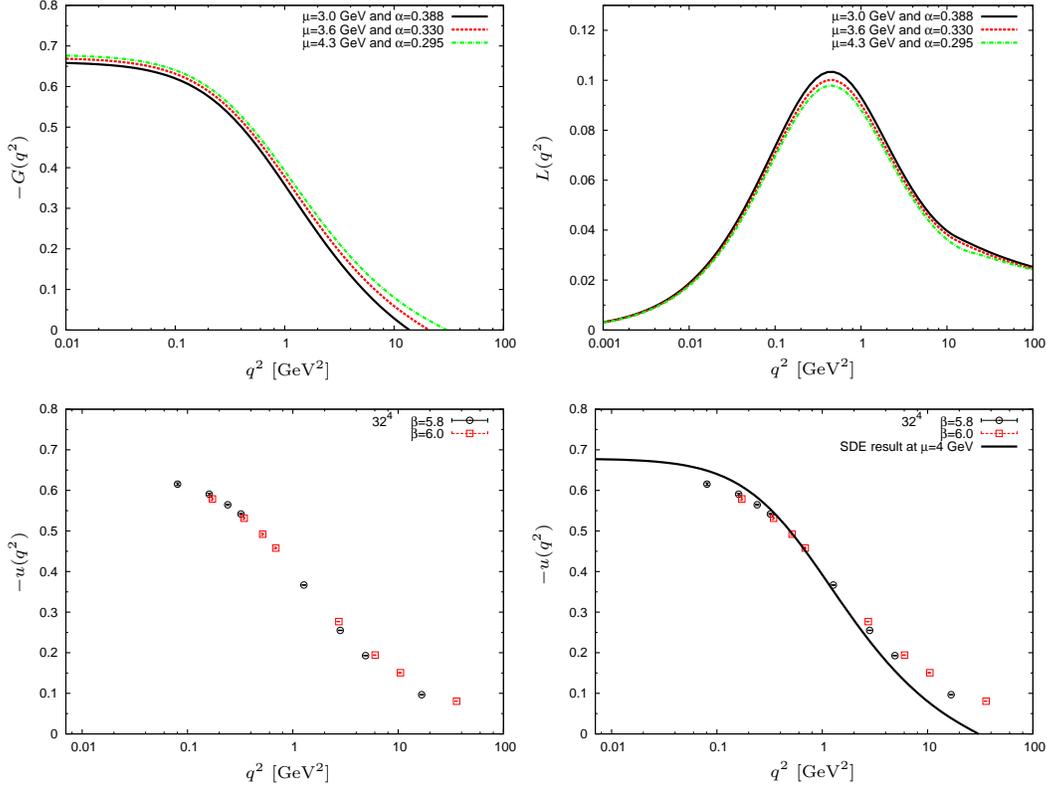}
\caption{The form factors $-G(q^2)$ and $L(q^2)$ determined from Eq. (4.2) at different normalization points $\mu$, using the procedure specified in the text. In the bottom panels we show the direct calculation on lattice of the Kugo-Ojima function $u(q^2)$ at $\mu=4$ GeV and a comparison with our result.}
\label{fig6}
\end{center}
\end{figure}

In Fig.~(\ref{fig6}) we show the auxiliary functions (\ref{simple}) calculated in $d=4$ at different renormalization points. One can see that indeed $L(0)=0$ and that in general $L(q^2)$ is suppressed with respect to $G(q^2)$ (also, power counting shows that $L$ is UV finite, while $G$ is not). In addition one finds that the function $G(q^2)$ saturates at an IR value bigger than $-1$ (around $-2/3$ for the renormalization points chosen) once again excluding IR enhancement of the ghost dressing function.
These results can be compared with the direct lattice calculations of the Kugo-Ojima function in terms of Monte Carlo averages~\cite{Sternbeck:2006rd}, though the comparison is at most suggestive because of systematic errors due to the different renormalization procedures employed.

\subsection{The effective charge}

Another important information that can be extracted from the PT-BFM equations is the running of the renormalization group (RG) invariant QCD effective charge, and, in particular, its behavior and value in the deep IR region. This quantity lies at the interface between perturbative and non-perturbative effects in QCD, providing a continuous interpolation between two physically distinct regimes: the deep UV, where perturbation theory is reliable, and the deep IR, where non-perturbative techniques must be employed. 

There are two possible RG invariant products  on which a definition of the effective charge can be based: $\widehat{r}(q^2)$ which exploits the non-renormalization property of the ghost vertex in the Landau gauge, and $\widehat{d}(q^2)$ which exploits the fact that BFM quantities  satisfy Ward (as opposed to Slavnov-Taylor) identities. One has
\be
\widehat{r}(q^2)=g^2(\mu^2)\Delta(q^2)F^2(q^2) \qquad \widehat{d}(q^2)=g^2(\mu^2)\widehat{\Delta}(q^2).
\label{RGIs}
\ee
These two {\it dimensionful} quantities, that have a mass dimension of $-2$, share an important common ingredient, namely the scalar cofactor of the gluon propagator $\Delta(q^2)$ which actually sets the scale. The next step is then to extract a {\it dimensionless} quantity that would correspond to the nonperturbative effective charge. Perturbatively, i.e., for asymptotically large momenta, it is clear that the mass scale is saturated simply by $q^2$, the bare gluon propagator, and the effective charge is defined by pulling a $q^{-2}$ out of the corresponding RG-invariant quantity. Of course,  in the IR the gluon propagator becomes effectively massive; therefore, particular care is needed in deciding exactly what combination of mass scales ought to be pulled out. The correct procedure in such a case~\cite{Cornwall:1981zr} is to pull out a massive propagator of the form (in Euclidean space) $[q^2+m^2(q^2)]^{-1}$, with $m^2(q^2)$ the dynamical gluon mass\footnote{Given that the gluon propagator is finite in the IR, if one insists on factoring out a simple $q^{-2}$ term, one would get a completely unphysical coupling, namely, one that vanishes in the deep IR, where QCD is expected to be (and is) strongly coupled.}.
One then has, making use of the background quantum identity~(\ref{BQI})
\be
\alpha_{\mathrm{gh}}(q^2)=\alpha(\mu^2)[q^2+m^2(q^2)]\Delta(q^2)F^2(q^2);\qquad 
\alpha(q^2)=\alpha(\mu^2)[q^2+m^2(q^2)]\frac{\Delta(q^2)}{[1+G(q^2)]^2}.
\label{echs}
\ee
In addition, due to the identity~(\ref{id}), the two effective charges are related through~\cite{Aguilar:2008fh}
\be
\alpha(q^2)=\alpha_{\mathrm{gh}}(q^2)\left[1+\frac{L(q^2)}{1+G(q^2)}\right]^2.
\ee
Since $L(0)=0$, we therefore see that not only the two effective charges coincide in the UV region where they should reproduce the perturbative result, but also in the deep IR where one has $\alpha(0)=\alpha_{\mathrm{gh}}(0)$. In addition, due to the relative suppression of the $L(q^2)$ form factor as compared to $G(q^2)$  even in the region of intermediate momenta~(Fig.~\ref{fig6}), where the difference reaches its maximum, the relative difference between the two charges is small (at most 10\%), making them practically  indistinguishable.

\begin{figure}[!t]
\begin{center}
\includegraphics[scale=1.1]{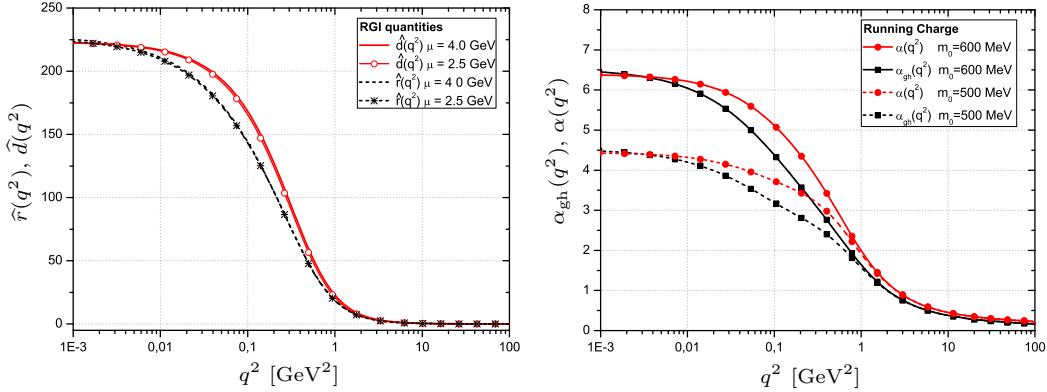}
\end{center}
\caption{\label{fig7}
{\it Left panel}: Comparison between the two RG-invariant products $\widehat{d}(q^2)$ (solid line) and $\widehat{r}(q^2)$ (dashed line); notice that there are two overlapping curves at different $\mu$ for each product. {\it Right panel}: 
Comparison between the QCD effective charge extracted from lattice data:  $\alpha(q^2)$ (red line with circles) 
and $\alpha_{\mathrm{gh}}$ (black line with squares) for two different masses: $m_0=500$ MeV (dashed) and $m_0=600$ MeV (solid).}
\end{figure}

In Fig.~\ref{fig7} we show both a check of the RG-invariance of the combinations~(\ref{RGIs}) as well as a comparison between the effective charges~(\ref{echs}) extracted from the lattice data for two different values of the running gluon mass~\cite{Aguilar:2008fh}). 

\section{Conclusions and outlook}

In this talk we have reviewed the PT-BFM framework for formulating (and solving) the SDEs of Yang-Mills theories. 
In particular, we have sketched how a dynamically generated gluon mass gives rise self-consistently to condensation of (thick) vortices leading to confinement, and thus that, contrary to naive expectations, {\it a gluon mass does not conflict with confinement, but enables it}.
In addition, we have derived within the PT-BFM framework a new set of SDEs that (i) accommodate the dynamical gluon mass generation through the Schwinger mechanism, and (ii) have  much better truncation properties than the conventional equations. The solutions of this new set of SDEs for the cases of the gluon propagator and the ghost dressing function shows IR saturation in agreement with {\it all} lattice simulations up to date.
The analysis of the effective charge has also revealed that {\it the gluon mass keeps the theory strongly coupled and allows for the presence of a conformal window}.

It should be emphasized that within this framework, and contrary to other approaches, the BRST symmetry of the theory remains intact throughout.  Indeed, at no point have we tampered with the  original Yang-Mills Lagrangian (and therefore we have no tree-level BRST violation), while the special transversality properties of the new SDEs ensure the preservation of the STIs (a direct consequence of the BRST symmetry),  and most importantly of the transversality condition $q^\nu\Pi_{\mu\nu} =0$,  at every step of the truncation. 

There are two complementary aspects that need to be further investigated. On the one hand, one would like to improve the agreement between the PT-BFM and lattice results in $d=4$. This is in our opinion linked to devising better vertex Ans\"atze 
for implementing the Schwinger mechanism. On the other hand, the lattice should address the issue of how 
the results changes when calculations are performed in gauges other than the Landau (recently an algorithm implementing $R_\xi$-type of gauges on the lattice has been developed, see~\cite{Cucchieri:2009kk}). In particular, a lattice implementation 
(along the lines suggested in~\cite{Dashen:1980vm}
of the BFM in the Feynman gauge, where all results are known to be free from gauge artifacts~\cite{Cornwall:1981zr,Cornwall:1989gv,Binosi:2002ft}, would be more than welcome. 

Finally, it is now timely to focus also on phenomenological issues in order to complement the lattice findings with experimental 
evidence. For example in~\cite{Luna:2005nz} the influence of an IR dynamical gluon mass scale in the calculation of $pp$ and $p\bar p$ forward scattering quantities through a QCD-inspired eikonal model was addressed, finding good agreement with data for a typical gluon mass of the order of 500 MeV. Also for hybrids with an extra gluon it was shown in~\cite{Takahashi:2004rw} 
that the best fit to lattice data on three heavy quarks plus a gluon definitely favor a gluon mass of the same order.

It is evidently this interplay between theory, lattice simulations and experiments that in the long run will prove to be most fruitful in unraveling the full IR dynamics of Yang-Mills theories.\\

\noindent {\it Acknowledgments}: We would like to thank the organizers of LC2010 for providing such a stimulating environment and enjoyable workshop.

\end{document}